\documentclass[twocolumn,showpacs,prl,amsmath,amssymb]{revtex4}
\usepackage{graphicx}
\usepackage{dcolumn}
\usepackage{bm}
\begin{document}

\title{Phonon Linewidths and Electron Phonon Coupling in Nanotubes}
\author{Michele Lazzeri$^1$}
\email{lazzeri@impmc.jussieu.fr}
\author{S. Piscanec$^2$, Francesco Mauri$^1$}
\author{A.C. Ferrari$^2$}
\email{acf26@eng.cam.ac.uk}
\author{J. Robertson$^2$}
\affiliation{ $^1$Institut de
Min\'eralogie et de Physique des Milieux Condens\'es,
4 Place Jussieu, 75252, Paris cedex 05, France \\
$^2$Department of Engineering, University of Cambridge, Cambridge
CB2 1PZ, UK}
\date{\today}

\begin{abstract}
We prove that Electron-phonon coupling (EPC) is the major source
of broadening for the Raman G and G$^{-}$ peaks in graphite and
metallic nanotubes. This allows us to directly measure the
optical-phonon EPCs from the G and G$^{-}$ linewidths. The
experimental EPCs compare extremely well with those from density
functional theory. We show that the EPC explains the difference in
the Raman spectra of metallic and semiconducting nanotubes and
their dependence on tube diameter. We dismiss the common
assignment of the G$^{-}$ peak in metallic nanotubes to a Fano
resonance between phonons and plasmons. We assign the G$^{+}$ and
G$^{-}$ peaks to TO (tangential) and LO (axial) modes.

\end{abstract}

\maketitle

Electron-phonon coupling (EPC) is a key physical parameter in
nanotubes. Ballistic transport, superconductivity, excited state
dynamics, Raman spectra and phonon dispersions all fundamentally
depend on it. In particular, the optical phonons EPC are extremely
relevant since electron scattering by optical phonons sets the
ultimate limit to high field ballistic
transport~\cite{dekker00,park04,lazzeri05,perebeinos05,javey04}.
Furthermore, they play a key role in defining the phonon
dispersions~\cite{piscanec04} and the Raman spectra of metallic
and semiconducting single wall nanotubes (SWNT), which show
different features
~\cite{pimenta98,kataura99,rafailov,brown01,jorio02,jiang02,kempa,
maul03,book, telg05, krupke05,doorn05}. Several theoretical and
experimental investigations of acoustic phonons EPC have been
published (see, e.g., Ref.~\cite{hertel00}).  However, only
tight-binding calculations of optical phonons EPC were performed,
with contrasting
results~\cite{dekker00,park04,saito04,jiang05,pennington04,mahan03}.
More crucially, no direct measurement of optical phonons EPC has
been reported to validate these calculations.

In this Letter we prove that the optical phonons EPC are the major
source of broadening for the Raman G and G$^{-}$ peaks in graphite
and metallic SWNTs. We show that the experimental Raman linewidths
provide a direct EPC measurement. The EPC are also responsible for
the the G$^+$ and G$^-$ splitting in metallic SWNTs. This allows
us to unambiguously assign the G$^{+}$ and G$^{-}$ peaks to TO
(tangential) and LO (axial) modes, in contrast to what often
done~\cite{pimenta98,kataura99,brown01,jorio02,jiang02,book}. We
dismiss the common assignment of the G$^{-}$ peak in metallic
SWNTs to a Fano resonance between phonons and
plasmons~\cite{brown01,jorio02,jiang02,paillet05,kempa,bose}.

In a perfect crystal, the linewidth $\gamma$ of a phonon is
determined by its interaction with other elementary excitations.
Usually, $\gamma = \gamma^{an}+\gamma^{EP}$, where $\gamma^{an}$
is due to the interaction with other phonons and $\gamma^{EP}$
with electron-hole pairs. $\gamma^{an}$ is determined by
anharmonic terms in the interatomic potential and is always there.
$\gamma^{EP}$ is determined by the EPC and is present only in
systems where the electron gap is zero.  If the anharmonic
contribution $\gamma^{an}$ is negligible or otherwise known,
measuring the linewidth is the simplest way to determine the EPC.

A phonon is described by a wave vector $\bf q$, branch index
$\eta$ and frequency $\omega_{{\bf q}\eta}$. We consider a
mean-field single particle formalism, such as density functional
theory (DFT) or Hartree-Fock. The EPC contribution to
$\gamma_{{\bf q}\eta}$ is given by the Fermi golden
rule~\cite{allen}:
\begin{eqnarray}
\gamma^{EP}_{{\bf q}\eta} &=& \frac{4\pi}{ N_{\rm k}}
\sum_{{\bf k},i,j} |g_{({\bf k}+{\bf q})j,{\bf k}i}|^2
[f_{{\bf k}i}-f_{({\bf k}+{\bf q})j}] \times \nonumber \\
&&\delta[\epsilon_{{\bf k}i} - \epsilon_{({\bf k}+{\bf q})j}
+ \hbar\omega_{ {\bf q}\eta}],
\label{eq1}
\end{eqnarray}
where the sum is on the electron vectors ${\bf k}$ and bands $i$
and $j$, $N_{\rm k}$ is the number of ${\bf k}$ vectors, $f_{{\bf
k}i}$ is the occupation of the electron state $|{\bf k},i\rangle$,
with energy $\epsilon_{{\bf k}i}$, $\delta$ is the Dirac
distribution. $g_{({ \bf k}+{\bf q})j,{ \bf k}i} = D_{({ \bf k}+{
\bf q})j,{ \bf k}i} \sqrt{\hbar/(2M\omega_{{\bf q}\eta})},$ where
$M$ is the atomic mass. $D_{({\bf k}+{\bf q})j,{\bf k}i}= \langle
{\bf k}+{\bf q},j| \Delta V_{{\bf q\eta}} |{\bf k},i\rangle$ and
$\Delta V_{{\bf q}\eta}$ is the potential derivative with respect
to the phonon displacement. $D$ is the EPC.

The electron states contributing to the sum in Eq.~\ref{eq1} are
selected by the energy conservation condition $\epsilon_{{\bf k}i}
+ \hbar\omega_{ {\bf q}\eta} = \epsilon_{({\bf k}+{\bf q})j}$.
Also, the state ${\bf k}i$ has to be occupied and $({\bf k}+{\bf
q})j$ empty, so that the term $[f_{{\bf k}i}-f_{({\bf k}+{\bf
q})j}] \ne 0$.  Thus, only electrons in the vicinity of the Fermi
level contribute to $\gamma^{EP}$. In insulating and
semiconducting systems $\gamma^{EP}=0$. In general, a precise
estimate of $\gamma^{EP}$ from Eq.~\ref{eq1} is possible only
after an accurate determination of the Fermi surface. However,
graphite and SWNTs are very fortunate cases. Thanks to their
particular band structure, $\gamma^{EP}$ is given by a simple
analytic formula.

In general, the EPC determines both phonon dispersions and
linewidths. We first consider the case of graphite and show that
both dispersion and linewidths give a direct measure of the EPC at
${\bm \Gamma}$.

The electron bands of graphite are well described by those of a two
dimensional graphene sheet. In graphene, the gap is zero for the
$\pi$ bands at the two equivalent ${\bf K}$ and ${\bf K'}=2{\bf
K}$ points of the Brillouin zone. We define $\langle D^2_{\bm
\Gamma}\rangle_F = \sum_{i,j}^\pi |D_{{\rm \bf K}i,{\rm \bf
K}j}|^2/4$, where the sum is on the two degenerate $\pi$ bands at
the Fermi level $\epsilon_F$. We consider the EPC relative to the
E$_{2g}$ phonon at ${\bm \Gamma}$. The $\Gamma$-E$_{2g}$ mode is
doubly degenerate and consists of an antiphase in-plane motion.
For a small non zero ${\bf q}$ near ${\bm \Gamma}$, this splits
into a quasi longitudinal (LO) and quasi transverse (TO) branch,
corresponding to an atomic motion parallel and perpendicular to
${\bf q}$. From DFT calculations we get $\langle D^2_{\bm
\Gamma}\rangle_F$~=45.60(eV/\AA)$^2$ for both LO and TO
modes~\cite{piscanec04}.

\begin{figure}
\centerline{\includegraphics[width=65mm]{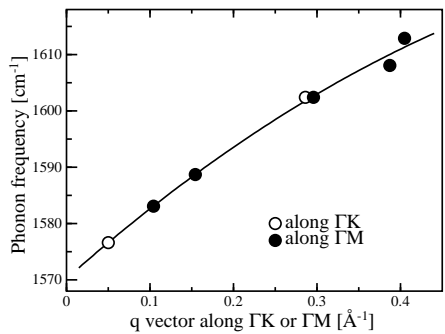}}
\caption{Graphite phonon dispersion of the highest optical branch
near ${\bm \Gamma}$. Dots are inelastic X-ray measurements from
Ref. ~\protect\cite{Maultzsch04}. The line is a quadratic fit.}
\label{fig1}
\end{figure}

We have previously shown that graphite has two Kohn anomalies in
the phonon dispersions for the ${\Gamma}$-E$_{2g}$ and ${\bf
K}$-A$'_1$ modes~\cite{piscanec04}. Due to the anomaly, the
dispersion near ${\bm \Gamma}$ of the E$_{2g}$-LO mode is almost
linear, with slope $S^{\rm LO}_{\bm \Gamma}$~\cite{piscanec04}:
\begin{equation}
S^{\rm LO}_{\bm \Gamma} =
\frac{\sqrt{3}\hbar a_0^2}{8M\omega_{\bm \Gamma}\beta}
\langle D^2_{\bm \Gamma}\rangle_F,
\label{eq2}
\end{equation}
where $a_0$=2.46~\AA~is the graphite lattice spacing,
$\beta$~=5.52~\AA~eV is the slope of the electron bands near
$\epsilon_F$, $M$ is the carbon atomic mass and $\omega_{\bm
\Gamma}$ is the frequency of the E$_{2g}$ phonon
($\hbar\omega_{\bm \Gamma}$=196.0 meV). Eq.~\ref{eq2} shows that
$\langle D^2_{\bm \Gamma}\rangle_F$ can be directly measured from
the experimental $S^{\rm LO}_{\bm \Gamma}$. The phonons around
${\bm \Gamma}$ have been measured by several groups with close
agreement. From a quadratic fit to the most recent data of
ref.~\cite{Maultzsch04}(Fig.~\ref{fig1}) we get $S^{\rm LO}_{\bm
\Gamma} = 133$~cm$^{-1}$\AA. From Eq.~\ref{eq2} we have $\langle
D^2_{\bm \Gamma}\rangle_F$ = 39~(eV/\AA)$^2$, in good agreement
with DFT.

\begin{figure}
\centerline{\includegraphics[width=85mm]{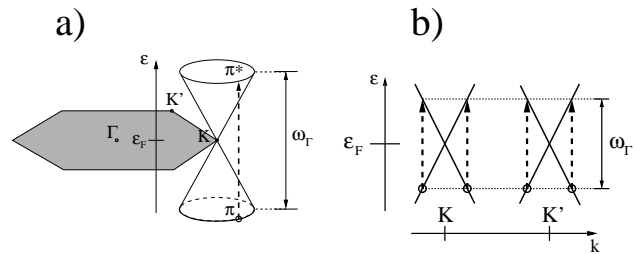}}
\caption{Electron bands around ${\bf K}$ in graphene ($a$) and in
a metallic tube ($b$). Shaded area is the graphene Brillouin zone.
Dashed arrows: decay processes for a ${\bm \Gamma}$ phonon.}
\label{fig2}
\end{figure}

We now show that the Full Width at Half Maximum (FWHM) of the
graphite G peak gives another independent EPC measurement. The G
peak of graphite is due to the ${\bm \Gamma}$-E$_{2g}$
phonon~\cite{Tuinstra}. We use Eq.~\ref{eq1} to compute the width,
$\gamma^{EP}_{\bm \Gamma}$, for this mode. Close to ${\bf K}$, we
assume the $\pi$ bands dispersion to be conic from the Fermi level
$\epsilon_{\rm F}$, with slope $\beta$
(Fig.~\ref{fig2}a)~\cite{nota00}. For both LO and TO modes:
\begin{equation}
\gamma^{EP}_{{\bm \Gamma}} =
\frac{\sqrt{3}a_0^2\hbar^2}{4M\beta^2} \langle D^2_{\bm \Gamma}\rangle_F.
\label{eq3}
\end{equation}
We then measure FWHM(G) for a single-crystal graphite similar to
that of Ref.~\cite{Maultzsch04}. Its Raman spectrum does not show
any D peak (Reich in~\cite{book} and ~\cite{scardaci05}), thus we
exclude extra broadening due to disorder~\cite{acf00}. By fitting
the G peak with a Lorentzian we get FWHM(G)=13 cm$^{-1}$.
Temperature dependent measurements show no increase of FWHM(G) in
the 2K-900K range~\cite{scardaci05}. Our Raman spectrometer
resolution is $\sim$1.5cm$^{-1}$~\cite{scardaci05}. We thus assume
an anharmonic contribution lower than the spectral resolution
$\gamma^{an}_G<1.5$cm$^{-1}$. Thus, we estimate $\gamma^{EP}_{\bm
\Gamma} \sim 11.5 $ cm$^{-1}$. Then, from Eq.~\ref{eq3},$\langle
D^2_{\bm \Gamma}\rangle_F\sim 47$~(eV/\AA)$^2$. This compares very
well with DFT, again confirming that $\gamma^{an}_G $ is small.

Finally, near ${\bm \Gamma}$ the conservation of the energy and
momentum in Eq.~\ref{eq1}, implies:
\begin{equation}
\gamma^{EP}_{\bf q} = 0 ~~\Leftrightarrow~~
{\rm q} \ge {\hbar \omega_{\bm \Gamma}}/{\beta}.
\label{eq4}
\end{equation}
This condition is satisfied by the E$_{2g}$ phonon, involved in the
Raman G peak
of graphite. On the other hand, the double resonant mode close to
${\bm \Gamma}$, which gives the D' peak at$\sim$1615 cm$^{-1}$,
does not satisfy it. D' is indeed sharper than the G
peak~\cite{Tan}.

In summary, we presented two independent measurements
of the graphite ${\bm \Gamma}$-E$_{2g}$ EPC. These are consistent with each other
and with DFT. We now consider SWNTs.

The G peak of SWNTs can be fit with only two components, G$^{+}$
and G$^{-}$~\cite{jorio02}. Semiconducting SWNTs have sharp
G$^{+}$ and G$^{-}$, whilst metallic SWNTs have a broad
downshifted
G$^{-}$~\cite{pimenta98,kataura99,rafailov,brown01,jorio02,jiang02,kempa,
maul03,book,telg05,krupke05,doorn05}. The G$^{-}$ band shows a
strong diameter dependence, being lower in frequency for smaller
diameters~\cite{jorio02}. This suggested its attribution to a
tangential mode, whose circumferential atoms displacements would
be most affected by a variation in diameter~\cite{brown01}. Thus,
the G$^{+}$ and G$^{-}$ peaks are often assigned to LO (axial) and
TO (tangential)
modes~\cite{pimenta98,kataura99,brown01,jorio02,jiang02,book}.

Conflicting reports exist on the presence and relative intensity
of the G$^{-}$ band in isolated versus bundled metallic tubes. It
has been claimed that this peak is as intense in isolated SWNTs as
in bundles~\cite{jorio02,krupke05,doorn05}; that it is
smaller~\cite{jiang02,maul03,telg05}; or that it can be
absent~\cite{paillet05}. The G$^{-}$ peak is also thought to
represent a Fano resonance due to phonon coupling with
plasmons~\cite{kempa,jiang02,brown01,bose,paillet05}. Such
phonon-plasmon coupling would either need~\cite{kempa} or not
need~\cite{brown01, bose} a finite phonon wavevector for its
activation. The theory of Refs.~\cite{bose,brown01} predicts the
phonon-plasmon peak to be intrinsic in single SWNT, in contrast
with~\cite{paillet05}. On the other hand, the theory in
Ref.~\cite{kempa} requires several tubes ($>$20) in a bundle in
order to observe a significant G$^{-}$ intensity, in contrast with
the experimental observation that bundles with very few metallic
tubes show a significant
G$^{-}$~\cite{maul03,telg05,krupke05,paillet05,doorn05}.
Ref.~\cite{kempa} also predicts a G$^{-}$ upshift with number of
tubes in the bundle, in contrast with Ref.~\cite{paillet05}, which
shows a downshift, and with Refs.~\cite{jorio02,brown01,krupke05},
which show that the G$^{-}$ position depends on the tube diameter
and not bundle size. Finally, the G$^{-}$ position predicted
by~\cite{kempa,bose} is at least 200 cm$^{-1}$ lower than that
measured
~\cite{pimenta98,kataura99,rafailov,brown01,jorio02,jiang02,
maul03,book,telg05,krupke05,doorn05}. Thus, all the proposed
theories for phonon-plasmon coupling~\cite{kempa,bose,brown01} are
very qualitative, require the guess of several physical
quantities, and fail to predict in a precise, quantitative,
parameter-free way the observed line-shapes and their dependence
on the SWNT diameter.
\begin{figure}
\centerline{\includegraphics[width=75mm]{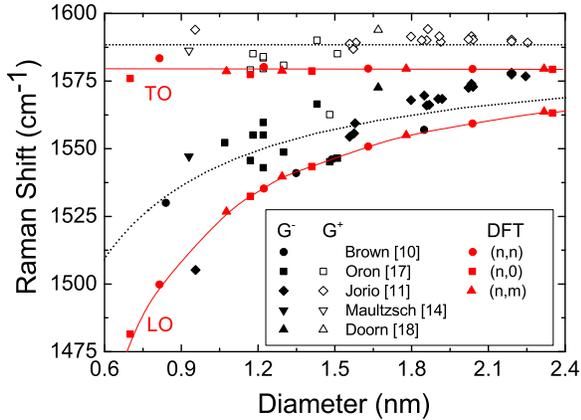}}
\caption{(color online) Black points: experimental G$^{+}$ and G$^{-}$
in metallic SWNTs vs. diameter.
Red points: DFT for armchair [from
(3,3) to (18,18)], zigzag [from (9,0) to (30,0)] and chiral tubes [
(5,2), (12,3), (16,1), (16,10), (20,14)].
Lines: fit of Eq. 5 to the experimental and DFT data.} \label{fig3}
\end{figure}

\begin{figure}
\centerline{\includegraphics[width=65mm]{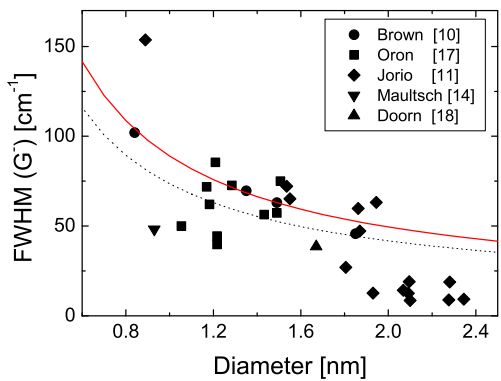}}
\caption{Experimental FWHM(G$^-$) in metallic SWNTs vs. diameter.
Solid line: DFT. Dotted line: fit of FWHM=$\gamma^{EP}_{{\bm
\Gamma}-{\rm LO}}+\gamma^{an}$, with $\gamma^{EP}_{{\bm
\Gamma}-{\rm LO}}$ from Eq.~\ref{eq5} and
$\gamma^{an}=10cm^{-1}$.} \label{fig4}
\end{figure}

We now show that, like in graphite, the EPC {\it per-se} gives the
main contribution to the G$^{-}$ position and FWHM, even in the
absence of phonon-plasmon coupling. Surprisingly, this has not
been considered so far. However, it is clear that, if phonons do
not couple to electrons, they certainly cannot couple to plasmons.

We compute the phonon frequencies for several metallic SWNTs using
DFT (details are in Refs.~\cite{lazzeri05,piscanec05}).  For all
metallic tubes, of {\it any chirality}, we get a splitting of the modes
corresponding to the graphite ${\bm \Gamma}$-E$_{2g}$ into quasi
transverse (TO) and a quasi longitudinal (LO), corresponding to an
atomic motion tangential to the SWNT axis or parallel to it. The
LO frequency is \emph{smaller} than the TO and has a strong
diameter dependence(Fig.~\ref{fig3}). This is the opposite of what
originally proposed, i.e. that a tangential mode should have a
much stronger diameter dependence than an axial
one~\cite{pimenta98,kataura99,brown01,jorio02,jiang02,book}. This
counter-intuitive result can be understood only considering the
presence of a Kohn anomaly in the phonon dispersion of metallic
SWNTs~\cite{piscanec04,Dubay,piscanec05}. The Kohn anomaly lowers
the frequency of phonons having a significant EPC. While the LO
EPC is large, the TO EPC is zero. Thus, the LO frequency is
smaller than the TO. Furthermore, from the dependence of the EPC on the
tube diameter $d$~\cite{lazzeri05}, we
get the analytic result~\cite{piscanec05}:
\begin{equation}
\omega^2_{LO}=\omega^2_{TO}-C_{M}/d \label{eq5a}
\end{equation}
with $C_{M}$ a constant. We thus assign the G$^-$ peak to the LO
(axial) mode and the G$^+$ to the TO (tangential) mode.
Fig.~\ref{fig3} compares our DFT calculations with experiments.
The agreement is excellent, considering that no fitting parameters
are used. If we fit Eq.~\ref{eq5a} to Fig.~\ref{fig3}, we get
$C_{M}$=1.52x10$^{5}$ cm$^{-2}$ nm; $\omega_{TO}$=1591 cm$^{-1}$.
Note that a 1/d$^{4}$ dependence of $\omega^2_{G^{-}}$ was
previously suggested based on phonon-plasmon
coupling~\cite{brown01,jorio02}.

Now we use Eq.~\ref{eq1} to derive the EPC contribution to
FWHM(G$^+$) and FWHM(G$^-$) in metallic SWNTs. The EPC of a SWNT
can be obtained from the graphite EPC $\langle D^2_{\bm
\Gamma}\rangle_F$ via zone-folding (valid for $d \geq$0.8 nm, i.e.
for the vast majority of SWNTs used in experiments and
devices)~\cite{lazzeri05}. Four scattering processes are involved
(Fig.~\ref{fig2}-b) and the LO and TO linewidths are:
\begin{equation}
\gamma^{EP}_{{\bm \Gamma}-{\rm LO}} =
\frac{2\sqrt{3}\hbar a_0^2}{\pi M \omega_{\bm \Gamma}\beta}
\frac{\langle D^2_{\bm \Gamma}\rangle_F}{d} ;
~~\gamma^{EP}_{{\bm \Gamma}-{\rm TO}}=0. \label{eq5}
\end{equation}
Eq.~\ref{eq5} is a key result. It shows that the EPC contributes
to the linewidth only for the LO mode in metallic SWNTs. For
semiconducting SWNTs the EPC contribution is zero for both the TO
and LO modes, since the gap does not allow to satisfy the energy
conservation in Eq.~\ref{eq1}.

This confirms our assignment of the G$^-$ and G$^+$ Raman peaks to
the LO and TO modes. Experimentally, in semiconducting SWNTs
FWHM(G$^+$) and FWHM(G$^-$) are similar (both are $\simeq
10$~cm$^{-1}$), while in metallic tubes
FWHM(G$^-$)$\simeq60~cm^{-1}$ and FWHM(G$^+$)$\simeq10$~cm$^{-1}$,
for a given $d$~\cite{jorio02,krupke05}. The large FWHM(G$^-$) in
metallic SWNTs is due the large $\gamma^{EP}$ and the small
FWHM(G$^+$) is entirely anharmonic in origin~\cite{nota02}. Thus,
even if FWHM(G) in graphite and FWHM(G$^+$) in SWNTs are similar,
their origin is totally different. This also explains why a
FWHM(G$^+$)$<$FWHM(G) can be seen~\cite{jorio02}.

Eq.~\ref{eq5} explains the 1/d dependence of FWHM(G$^-$) in
metallic SWNTs~\cite{jorio02}.  Then, using Eq.~\ref{eq5},
$\langle D^2_{\bm \Gamma}\rangle_F$ can be directly fit from the
experimental FWHM(G$^-$) (Fig.~\ref{fig4}).  We find $\langle
D^2_{\bm \Gamma}\rangle_F=37$(eV/\AA)$^2$, again in agreement with
DFT. Note that, while the DFT $\langle D^2_{\bm \Gamma}\rangle_F$
is that of planar graphene, the fitted $\langle D^2_{\bm
\Gamma}\rangle_F$ is obtained from measurements of SWNTs. Thus,
the good agreement between the two values is a direct verification
that the SWNT EPC can be obtained by folding the graphene
EPC~\cite{lazzeri05}.

The most common process involved in Raman scattering is single
resonance. Double resonance is necessary to explain the activation
of otherwise inactive phonons away from $\Gamma$, such as the D
peak ~\cite{Thomsen00,maul03}. It has been suggested that even the
G$^{+}$ and G$^{-}$ peaks in SWNTs are always double
resonant~\cite{Maul02,book}. However, if the laser excitation
energy satisfies single resonance conditions, the intensity of
single resonance peaks is expected to be dominant in the Raman
spectrum~\cite{pimenta03}. Double resonance can only be relevant
for higher excitation energies. The condition set by Eq.~\ref{eq4}
must also hold for SWNTs in order to have a significant EPC
contribution to the linewidth. Eq.~\ref{eq4} can only be satisfied
by phonons with wavevector too small to be double
resonant~\cite{book,maul03}. Thus, in double resonance, the
G$^{-}$ peak should be much narrower than experimentally observed.
Interestingly, it has been reported that the broad G$^-$ peak
disappears by increasing the excitation energy while measuring the
Raman spectrum of a metallic SWNT, and a sharper one
appears~\cite{maul03}. We explain this as a transition from single
to double resonance.

Finally, in Ref.~\cite{lazzeri05} we computed the zero-temperature
life-time $\tau$ of a conduction electron in a SWNT, due to phonon
scattering.  If phonons are thermalized, $\tau$ determines the electron
mean free path $l_0=\beta\tau/\hbar$ at zero temperature.  The
phonon linewidths can be expressed as a function of $\tau$ as:
\begin{equation}
\gamma^{EP}_{{\bm \Gamma}-LO}= 4/{\tau^{bs}_{{\bm \Gamma}-LO}} =
4/{\tau^{fs}_{{\bm \Gamma}-TO}}, \label{eq6}
\end{equation}
where $bs$($fs$) indicate back
(forward)-scattering~\cite{lazzeri05}.  Such simple relations
might seem surprising. In fact, the two quantities $\gamma^{-1}$
and $\tau$ describe two distinct phenomena: $\gamma^{-1}$ is the
life-time of a phonon (Eq.~\ref{eq1}) and $\tau$ is the life-time
of an electron (Eq.~1 of Ref.~\cite{lazzeri05}). In general, one
does not expect a simple relation between $\gamma^{-1}$ and
$\tau$. However, due to the low dimensionality, in SWNTs
$\gamma^{EP}$ determines directly $\tau$ (Eq.~\ref{eq6}).  Thus,
the measured $\gamma^{EP}$ are also a direct measurement of $l_0$
for metallic SWNTs.  Since the measured $\gamma^{EP}$ is in
agreement with DFT calculations, the results of
Ref.~\cite{lazzeri05} are further confirmed.

In conclusion, we presented a set of simple formulas
(Eqs.~\ref{eq2}, \ref{eq3}, \ref{eq5}), with the graphite EPC as
the \emph{only} fit parameter. Remarkably, these formulas
\textit{quantitatively} explain a string of experiments, ranging
from phonon slopes and FWHM(G) in graphite to ${G^-}$ peak
position and FWHM(G$^{-}$) diameter dependence in SWNTs. Fitting a
wide set of independent data we obtain the \textit{same} EPCs
$\pm$10$\%$. These experimental EPC are in excellent agreement
with, and validate, the DFT approach.

Calculations performed at HPCF (Cambridge) and IDRIS (grant
051202).We acknowledge S. Reich and C. Thomsen for providing
single crystal graphite; M. Oron and R. Krupke for useful
discussions. Funding from EU IHP-HPMT-CT-2000-00209 and CANAPE,
from EPSRC GR/S97613 and The Royal Society is acknowledged.


\begin{thebibliography}{99}

\bibitem{dekker00}
Z. Yao, C.L. Kane and C. Dekker,
Phys. Rev. Lett. {\bf 84}, 2941 (2000).

\bibitem{park04}
J. Y. Park {\it et al.},
Nano Lett. {\bf 4}, 517 (2004).

\bibitem{lazzeri05}
M. Lazzeri et al. cond-mat/0503278.

\bibitem{perebeinos05}
V. Perebeinos, J. Tersoff, P. Avouris,
Phys. Rev. Lett.{\bf 94}, 086802 (2005).

\bibitem{javey04}
A. Javey {\it et al.},
Phys. Rev. Lett. {\bf 92}, 106804 (2004).

\bibitem{piscanec04}
S. Piscanec et al.
Phys. Rev. Lett. {\bf 93}, 185503 (2004).

\bibitem{pimenta98}
M.A. Pimenta {\it et al.},
Phys. Rev. B {\bf 58}, 16016 (1998).

\bibitem{kataura99}
H. Kataura {\it et al.}, Synth. Metals {\bf 103}, 2555 (1999).

\bibitem{rafailov}
P. M. Rafailov, H. Jantoljak, and C. Thomsen,
Phys. Rev. B {\bf 61}, 16179 (1999).

\bibitem{brown01}
S.D.M. Brown {\it et al.},
Phys. Rev. B {\bf 63}, 155414 (2001).

\bibitem{jorio02}
A. Jorio {\it et al.}, Phys. Rev. B {\bf 65}, 155412 (2002); {\bf
66}, 115411 (2002);M. S. Dresselhaus {\it et al.}, Physica B {\bf
323}, 15 (2002).

\bibitem{jiang02}
C. Jiang {\it et al.}, Phys. Rev. B {\bf 66}, 161404 (2002).

\bibitem{kempa}
K. Kempa, Phys. Rev. B {\bf 66}, 195406 (2002).

\bibitem{maul03}
J. Maultzsch et al. Phys. Rev. Lett. {\bf 91}, 087402 (2003).

\bibitem{book}
Jorio {\it et al.}; Thomsen {\it et al.}; Reich {\it et al.}; P.
Tan et al. in {\it Raman spectroscopy in carbons: from nanotubes
to diamond}, eds. A. C. Ferrari and J. Robertson Phil. Trans. Roy.
Soc. A {\bf 362}, 2267-2565 (2004).

\bibitem{telg05}
H. Telg {\it et al.}, {\it Proceedings of IWEPNM 2005} AIP, Melville, NY,
2005.

\bibitem{krupke05}
M. Oron, R. Krupke {\it et al.} Nano Lett. in press (2005).

\bibitem{doorn05}
S.K. Doorn {\it et al.}, Phys. Rev. Lett. {\bf 94}, 016802 (2005).

\bibitem{hertel00}
T. Hertel and G. Moos, Phys. Rev. Lett. {\bf 84}, 5002 (2000).

\bibitem{saito04} J. Jiang {\it et al.},
Chem. Phys. Lett. {\bf 392}, 383 (2004).

\bibitem{mahan03} G.D. Mahan, Phys. Rev. B {\bf 68}, 125409 (2003).

\bibitem{pennington04}
G. Pennington and N. Goldsman, Phys. Rev. B {\bf 68},045426 (2004).

\bibitem{jiang05}
J. Jiang {\it et al.}, Phys. Rev. B {\bf 71}, 045417 (2005).

\bibitem{paillet05}
M. Paillet {\it et al.}, Phys Rev. Lett. {\bf 64}, 327401 (2005).

\bibitem{bose}
S.M. Bose, S. Gayen, S.N. Behera, cond-mat/0506004.

\bibitem{allen}
P.B. Allen, Phys. Rev. B {\bf 6}, 2577 (1972);
P.B. Allen and R. Silberglitt, {\it ibid.} {\bf 9}, 4733 (1974).

\bibitem{Maultzsch04}
J. Maultzsch et al.
Phys. Rev. Lett. {\bf 92}, 075501 (2004).

\bibitem{Tuinstra}
F. Tuinstra, J.Koenig J. Chem. Phys. {\bf 53}, 1126 (1970).

\bibitem{nota00}
The EPC dependence on ${\bf k}$ is given in Eq.~6 of
Ref.~\protect\cite{piscanec04}. A factor 2 in
Eq.~\protect\ref{eq3} accounts for both ${\bf K}$ and ${\bf K'}$.

\bibitem{scardaci05}
V. Scardaci, A. C. Ferrari, unpublished.

\bibitem{acf00}
A.C. Ferrari and J. Robertson, Phys. Rev. B {\bf 61}, 14095 (2000);
{\it ibid.} {\bf 64}, 075414 (2001).


\bibitem{Tan}

P. Tan et al. Phys. Rev B {\bf 64}, 214301 (2000).

\bibitem{piscanec05}
S. Piscanec {\it et al.}, unpublished.

\bibitem{Dubay}
O. Dubay and G. Kresse, Phys. Rev. B {\bf 67}, 035401 (2003).

\bibitem{nota02}
FWHM(G$^-$) and FWHM(G$^+$) increase $\sim$15 cm$^{-1}$ with
temperature in the 80K-600K range for both metallic and
semiconducting SWNTs~\cite{jorio02,scardaci05}. Thus
$\gamma^{an}\ll\gamma^{EP}$.

\bibitem{Maul02}

J. Maultzsch, S. Reich, C. Thomsen, Phys. Rev. B {\bf 65} 233402
(2002)

\bibitem{Thomsen00}
C. Thomsen, S. Reich Phys. Rev. Lett. {\bf 85}, 5214 (2000).

\bibitem{pimenta03}

M.A. Pimenta {\it et al.}, in {\it Proceedings of IWEPNM 2003}
(AIP, Melville NY, 2003).

\end{thebibliography}
\end{document}